\documentclass[acmsmall]{acmart}

\usepackage[english]{babel}
\usepackage{blindtext}
\usepackage{tikz}
\usepackage{amsmath}

\usepackage{filecontents}
\usepackage{booktabs}
\usepackage{pifont}
\usepackage{enumitem}
\usepackage{xurl}
\usepackage{multirow}
\usepackage{graphicx} 
\usepackage{tabularx}
\usepackage{booktabs}
\usepackage{graphicx}
\usepackage{subcaption}
\usepackage{microtype}
\usepackage{float}
\usepackage[table]{xcolor}
\usepackage{caption}
\usepackage{makecell}

\definecolor{deepgreen}{rgb}{0.0, 0.5, 0.0}
\definecolor{deepred}{rgb}{0.7, 0.05, 0.0}
\newcommand{\cmark}{\textcolor{deepgreen}{\ding{51}}}
\newcommand{\xmark}{\textcolor{deepred}{\ding{55}}}
\newcolumntype{L}{>{\raggedright\arraybackslash}X}
\AtBeginDocument{%
  }

\setcopyright{none}
\renewcommand\footnotetextcopyrightpermission[1]{}  
\begin{document}

\title{One-Prompt Censorship Evasion via Generative Diffusion Models}

\author{Shiyi Ling}
\email{sling9@ucsc.edu}

\author{Yuhang Gan}
\email{ygan11@ucsc.edu}

\author{Chen Qian}
\email{cqian12@ucsc.edu}

\affiliation{%
  \institution{University of California, Santa Cruz}
  \city{Santa Cruz}
  \state{California}
  \country{USA}
}


\begin{abstract}
  The escalating arms race between Internet censorship and evasion has driven censors to evolve from static rule-based filtering to sophisticated deep learning-based traffic analysis. While recent automated evasion tools have attempted to counter this by leveraging stochastic search and programmable heuristics, they continue to suffer from insufficient evasion robustness across diverse censorship modalities and poor usability due to complex, mechanism-specific configurations that require manual fitness tuning or domain-specific languages. In this paper, we propose a paradigm shift that reframes censorship evasion as a semantic image-to-image editing task, allowing users to execute it with a single prompt. We introduce \emph{FlowPaint}, a novel generative framework that leverages the "world knowledge" of large diffusion models to automatically reshape censored traffic into benign patterns. \emph{FlowPaint} utilizes an instruction-tuned diffusion architecture to perform semantic editing on network flows. Evaluations against both industrial-grade rule-based middleboxes and learning-based classifiers demonstrate that \emph{FlowPaint} outperforms existing censorship evasion baselines, enabling users to counter diverse censorship paradigms solely by varying natural language instructions.
\end{abstract}

\ccsdesc[500]{Networks~Network security}
\ccsdesc[500]{Computing methodologies~Artificial intelligence}

\keywords{Censorship evasion, Diffusion models, Generative AI}


\maketitle

\section{Introduction}
\label{sec:intro}
Internet censorship has become a pervasive barrier to global connectivity, restricting the fundamental right to access and exchange information~\cite{Burnett2010-hg, ensafi2015examining}. To enforce these restrictions, censorship infrastructures typically deploy on-path middleboxes to identify and block prohibited traffic. These systems continuously monitor flows to extract specific features, ranging from explicit text signatures in unencrypted packets to statistical fingerprints in encrypted connections.

Censorship evasion has established itself as a persistent and critical domain within network security research. Beyond the technical challenge of circumventing restrictive censors, this field serves as a vital safeguard for digital human rights, ensuring the continuous availability of information in the face of restriction~\cite{Dixon2016-mj, ensafi2015examining}. As censors relentlessly upgrade their detection capabilities, evolving from simple keyword filtering to advanced Deep Packet Inspection (DPI) and, more recently, deep learning-based traffic analysis, as demonstrated in academic adversarial settings~\cite{Nasr2021-dj, Li2022-vl}, evasion technologies are necessitated to innovate continuously. Consequently, developing resilient and adaptive evasion mechanisms is essential to withstand increasingly sophisticated future censorship.

To counter evolving detection, traditional tools like Tor's Obfs4~\cite{obfs4} employ fixed cryptographic rules to randomize traffic fingerprints. While effective against static inspection, these rigid patterns remain vulnerable to statistical analysis. Consequently, systems such as \textit{Geneva}~\cite{geneva2019} and \textit{UPGen}~\cite{upgen} increasingly automate strategy discovery via evolutionary algorithms and procedural generation. 

Although these systems marked a significant leap forward, they face two persistent limitations: \textbf{insufficient evasion robustness} and \textbf{complex configurations}. First, these generative evasion strategies often overfit to specific censors, failing to maintain high success rates across heterogeneous adversaries: Obfs4 defeats signature-based DPI yet remains trivially identifiable by statistical classifiers. Second, these tools rely on complex, mechanism-specific configurations that require manual fitness tuning or domain-specific languages. For example, Geneva requires users to define genetic algorithm fitness functions over TCP mutations, while Marionette~\cite{Dyer2015-wh} demands hand-authored probabilistic format grammars. This complexity precludes out-of-the-box usability and severely hinders their deployment across varying network environments.

In this paper, we argue that overcoming these limitations requires a fundamental paradigm shift. Rather than viewing censorship evasion as a problem of manual protocol engineering or stochastic search, we reframe it as a \textbf{semantic traffic editing task}: reshaping a flow's statistical and header features to match the benign traffic manifold while preserving its encrypted payload. Inspired by the transformative success of Generative AI in computer vision, where models can edit complex images based on simple text prompts~\cite{Brooks2023-xa}, we ask: \textit{Can we leverage the "world knowledge" embedded in large generative models to "repaint" censored network flows automatically into benign forms?}

We answer this question with \emph{FlowPaint}, a novel generative framework that acts as a general flow editor for circumventing censorship. The core intuition behind \emph{FlowPaint} is that the packet headers of flows can be effectively mapped into the image domain, exploiting the spatial correlations inherent in byte streams. The visual representation paradigm has yielded advances in both traffic generation~\cite{Jiang2024-me} and intrusion detection~\cite{wang2024few, Shapira2021-FlowPic}. Motivated by this visual modality, we adapt the instruction-guided diffusion framework~\cite{sdxl_instructpix2pix_hf} to the censorship evasion domain. Unlike prior works that rely on random obfuscation or rule-based mimicry, \emph{FlowPaint} takes a censorship-triggering flow and a natural language instruction as input, iteratively refines the flow-based image using diffusion models, and converts the refined image to protocol-compatible packet headers. Specifically, it modifies \textit{statistical patterns} and \textit{optional header fields}, while leaving encrypted payloads intact; censors that analyze byte-level entropy or n-gram statistics of payload content are outside the scope of this work.

This generative approach has several significant advantages. First, it enhances \textbf{evasion robustness} by leveraging the expressive power of diffusion models. Instead of overfitting to specific rules, the diffusion model learns the latent distribution of benign traffic, generating high-fidelity patterns that bypass both rigid filters and adaptive classifiers. Second, it \textbf{minimizes operational complexity} through a natural language interface, enabling one-prompt evasion at inference time without manual configuration or domain-specific knowledge. Third, it achieves evasion through \textbf{in-place feature modification} without injecting dummy packets, minimizing bandwidth overhead.

This paper makes the following contributions:

\textbf{(1)} We are the first to reframe censorship evasion as a semantic image-to-image editing task, leveraging the generative power of diffusion models to automatically reshape censored flows into benign patterns under natural language guidance.

\textbf{(2)} We design and implement \emph{FlowPaint}, enabling 
one-prompt evasion at inference time without manual fitness functions, 
protocol specifications, or per-run training. Evaluations against learning-based classifiers and GFW-style middleboxes show that \emph{FlowPaint} achieves high evasion success rates where traditional tools fail, forcing censors into a dilemma of unacceptable over-blocking.

\textbf{We plan to open-source \emph{FlowPaint} upon acceptance.} The remainder of this paper is organized as follows. Section~\ref{sec:background} provides the background. Section~\ref{sec:method} introduces the design of \emph{FlowPaint}, and Section~\ref{sec:implementation} describes its implementation. The evaluation results are shown in Section~\ref{sec:evaluation}, followed by ablation studies in Section~\ref{sec:ablation}. Finally, we discuss some related issues in Section~\ref{sec:discussion}, review related work in Section~\ref{sec:related_work}, and conclude the paper in Section~\ref{sec:conclusion}.

\section{Background and Motivation}
\label{sec:background}

\subsection{Challenges of Censorship Evasion}
\label{sec:bg_evasion_challenges}
Driven by advancing censorship ~\cite{Dyer2013-up, Nourin2023-fo, Dixon2016-mj, Wright2009-mp, Wang2015-ea, Gong2020-hd}, evasion strategies have evolved from simple tunneling to complex traffic obfuscation~\cite{Wustrow2011-rt, Winter2013-ym, niere2025tls}. However, current methodologies face two fundamental limitations: insufficient robustness against censorship and complex configurations.

\noindent\textbf{Insufficient Evasion Robustness against Distinct Censors.}
The primary failing of existing tools is their tendency to specialize in a single evasion strategy while remaining vulnerable to others.  For instance, cryptographic obfuscators such as \textit{Obfs4} and \emph{ScrambleSuit}~\cite{Winter2013-ym} randomize payload entropy to defeat signature-based DPI. However, this very randomization creates a high-entropy statistical profile that makes them trivial targets for statistical traffic analysis~\cite{Wang2015-ea}. Adversarial patch methods designed to evade machine learning (ML) classifiers~\cite{Nasr2021-dj, Li2022-vl} often introduce header anomalies that, while confusing to a neural network, are immediately flagged by strict rule-based middleboxes. Even automated systems like \textit{UPGen}~\cite{upgen}, which attempt to mimic protocol structures, rely on rigid procedural generation logic. Lacking a deep semantic understanding of legitimate flow patterns, these systems struggle to adapt when adversaries employ both static filtering rules and dynamic analysis simultaneously.

\noindent\textbf{Rigid Configuration and Operational Complexity.}
A second critical barrier is the high level of domain expertise required to operate current evasion systems, which precludes widespread deployment. Evolutionary systems like \textit{Geneva}~\cite{geneva2019}, for example, rely on genetic algorithms that require users to define complex fitness functions and engage in resource-intensive training loops directly against the live censor. Similarly, programmable frameworks like \emph{Marionette}~\cite{Dyer2015-wh} demand that users manually write comprehensive probabilistic models for traffic formats. 
This complexity creates a bottleneck: human operators cannot rapidly configure or update countermeasures in response to abrupt shifts in censorship policies, rendering these tools impractical.

\subsection{Generative AI and Diffusion Models}
\label{sec:bg_diffusion}
In recent years, generative AI has demonstrated powerful modeling capabilities across diverse domains. Beyond their transformative impact on natural language processing, these technologies have shown significant potential in networking tasks, ranging from protocol fuzzing~\cite{Meng2024-jj} and automated network algorithm design~\cite{He2024-xv} to high-fidelity traffic generation~\cite{Jiang2024-me}. 

Diffusion Probabilistic Models (DDPM)~\cite{ho2020denoising} have recently emerged as the predominant framework for synthesizing high-fidelity data across various modalities. Diffusion models excel at learning complex data distributions by iteratively refining noise into structured data~\cite{rombach2022high}. Building on this, recent advances in computer vision have introduced the paradigm of \emph{instruction-guided semantic image editing}. Notable works such as \emph{InstructPix2Pix}~\cite{Brooks2023-xa} leverage the diffusion architecture to edit images via simple natural language prompts while preserving structural integrity.

We observe that censorship evasion is conceptually analogous to this diffusion-based image editing process. Transforming a censored flow into a benign one resembles replacing a “rose” with a “sunflower” in a photograph: the goal in both cases is to alter specific visual elements to match a target appearance while preserving the overall composition and style.

This work explores the adaptation of diffusion models to censorship evasion, suggesting that this framework is uniquely suited to address the challenges of robustness and usability. For robustness, unlike traditional obfuscators that rely on rigid heuristics, diffusion models can learn the overall statistical manifold of benign traffic while preserving the flow's underlying protocol structure. 
Simultaneously, regarding usability, text-conditioning mechanisms lower the entry barrier by mapping natural language instructions directly to low-level network 
parameters, replacing complex manual configurations with intuitive semantic control.

\section{FlowPaint Framework} 
\label{sec:method} 
Leveraging these generative strengths discussed above, we present \emph{FlowPaint}. In this section, we formulate the censorship evasion process as a semantic image-to-image editing task, detailing the design of \emph{FlowPaint}. 

\begin{figure}[t!]
    \centering
    \includegraphics[width=0.8\linewidth]{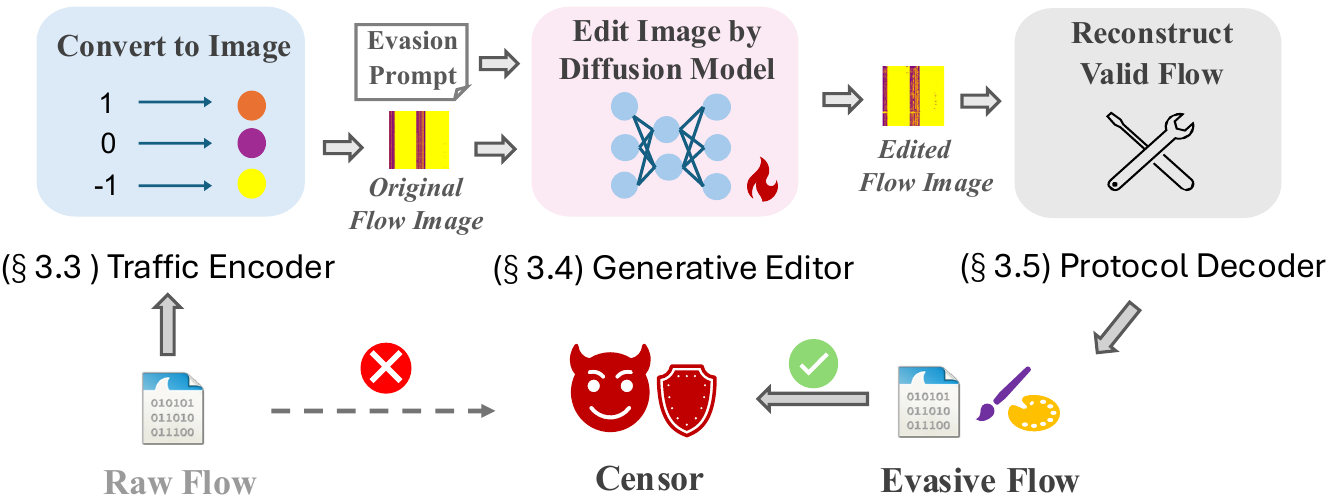} 
    \caption{\textbf{Work Flow of \emph{FlowPaint}.} Traffic Encoder transforms raw flows into images (\S\ref{sec:method_preprocess}). Generative Editor refines these representations (\S\ref{sec:method_diffusion}). Protocol Decoder projects the outputs back into valid traffic (\S\ref{sec:method_postprocess}).}
    \label{fig:architecture}
\end{figure}

\subsection{Overview}
\label{sec:method_overview}
\emph{FlowPaint} transforms censorship-triggering flows into benign patterns through instructions. 
The core intuition driving our approach is that network flows, when visualized as a bit-level bitmap, exhibit distinct textural patterns that differentiate benign traffic distributions from censorship-triggering flows. By leveraging the semantic understanding of the generative capabilities of diffusion models, \emph{FlowPaint} learns to "repaint" the feature space of a censored flow into a benign representation under natural language guidance.

\emph{FlowPaint} comprises three components, as illustrated in Figure~\ref{fig:architecture}. First, \textbf{Traffic Encoder} transforms discrete packet sequences into a standardized image suitable for diffusion models. Second, \textbf{Generative Editor} acts as the core engine, employing a fine-tuned diffusion model to "edit" the flow's features based on an evasion prompt while preserving the structural integrity of the original flow. Finally, \textbf{Protocol Decoder} projects the probabilistic output of the Generative Editor back onto the manifold of valid network protocols to produce protocol-compliant flows. Given a \textit{raw input flow} and a \textit{high-level prompt}, \emph{FlowPaint} automatically orchestrates the encoding, editing and reconstruction phases to output a valid flow that bypasses the specified censorship mechanism.


\subsection{Instruction-Guided Semantic Editing}
\label{sec:method_paradigm}
We formulate the problem of censorship evasion not as generation, but as instruction-guided semantic editing task. Standard text-to-image diffusion models (e.g., Stable Diffusion~\cite{rombach2022high}) generally synthesize novel content from Gaussian noise conditioned solely on text. While powerful, this paradigm is ill-suited for our task because it lacks control over the fidelity to the original input. In networking domain, generating a flow ``from scratch'' risks destroying the underlying protocol semantics required for valid communication.

To address this, we adopt the generative editing paradigm introduced by InstructPix2Pix~\cite{Brooks2023-xa}, which extends diffusion models to perform prompt-guided modifications on existing images. Unlike standard generation, this paradigm conditions the diffusion process on two inputs simultaneously: the source data state $x$ and a textual directive (or prompt) $c$. Formally, instead of predicting noise from a pure latent state, the model predicts a noise residual $\epsilon_\theta(x_t, t, x, c)$ that shifts the input distribution $x$ toward a target state $y$. 
This formulation decouples \textit{structural integrity} from \textit{stylistic attributes}: preserving global flow structure while reshaping the statistical features that trigger censors.
Specifically, the model can modify the flow's "texture," which encompasses \textit{statistical patterns} and \textit{optional header fields}, to evade detection. Simultaneously, it preserves the global "structure" necessary for connection stability. 

\subsection{Traffic Visualizing with Traffic Encoder}
\label{sec:method_preprocess}
We need to bridge the semantic gap between packet binaries and visual representations. However, raw network traffic consists of discrete, variable-length sequences, whereas diffusion models excel at processing continuous, fixed-dimensional spatial data. We address this challenge through a specialized three-stage encoding pipeline.

\noindent\textbf{Bit-level Matrix Construction}
\label{sec:method_preprocess_matrix}
The first stage addresses the challenge of creating a unified representation for variable-length network flows. We begin with raw traffic files (\texttt{.pcap}), but directly feeding raw byte streams into a generative model is suboptimal due to protocol misalignment. To resolve this, we adopt the standardized bit-level representation, \emph{nPrint}~\cite{nprintml}, to project the packet sequences into a fixed-width visual format. This process aligns packet headers against a unified protocol schema, encoding each bit position using a tri-state logic: active and inactive bits are represented as $1$ and $0$, respectively, while fields corresponding to missing headers are marked as $-1$. Ultimately, we construct a high-dimensional matrix where each row represents a single packet and columns correspond to standardized bit positions across the protocol stack (encompassing IPv4, TCP, UDP, and ICMP). To handle protocol variability, the matrix reserves static column indices for the maximum possible space of TCP and IP Options, ensuring a fixed-width representation.

We choose to use this \emph{nprint} representation because it constitutes a sufficient state space for our task. Since traffic payloads are encrypted by modern encrypted protocols, censorship systems predominantly rely on unencrypted metadata and side-channel features to identify blocked flows~\cite{Wang2015-ea, Li2022-vl}. Furthermore, by excluding payload data from the feature matrix, this representation strictly isolates protocol metadata from application content. This confines all generative modifications to the headers, ensuring the underlying encrypted payload remains intact and unmodified.

\noindent\textbf{Selective Masking for Generalization}
\label{sec:method_preprocess_masking}
While the \emph{nprint} matrix successfully aligns protocol structures, training the diffusion model directly on it introduces a critical risk: the model may learn to edit identity features such as source IP addresses, rather than the statistical patterns required to cheat a censor. This shortcut learning is catastrophic: IP addresses are non-editable routing identifiers, and altering them neither achieves evasion nor preserves network reachability.

To prevent the model from learning these identity shortcuts, we implement a selective masking strategy on the \emph{nPrint} matrix. We explicitly strip fields         containing high-entropy endpoint identities, such as source and destination IP addresses, blinding the model to routing information. We retain all remaining structural and stateful features, including TCP window size, TTL, IP total length (reflecting packet sizing), optional header fields such as NOP and MSS, and TCP flags and sequence-related fields, as these constitute the editable statistical fingerprint targeted by the diffusion model. By removing identity artifacts while preserving these shape-defining bits, we force the model to focus on the generalizable statistical patterns that characterize benign traffic, rather than overfitting to flow-specific identifiers.         

\noindent\textbf{Ternary Color Encoding and Image Normalization}
\label{sec:method_preprocess_color}
Following the selective masking, we obtain a refined feature matrix composed of three logical states. However, standard diffusion architectures are designed to ingest continuous, image data rather than discrete, single-channel sparse matrices. To bridge this gap, aligning with the NetDiffusion traffic-to-image representation approach ~\cite{Jiang2024-me}, we transform the \emph{nprint} matrix into a compatible visual format through \textit{Ternary Color Encoding}.

We project the discrete logical states into high-contrast chromatic values within the RGB space. Specifically, we map active bits ($1$) to orange, inactive bits ($0$) to purple, and placeholders ($-1$) to yellow. These three colors exhibit high mutual contrast in RGB space, making them well-suited for encoding a three-state discrete signal. This distinct chromatic separation serves a dual purpose. First, it resolves the semantic ambiguity between "value zero" and "data absence," allowing the convolutional layers to sharply distinguish between a flag that is explicitly turned off and a header field that effectively does not exist. Second, as visualized in Figure~\ref{fig:repaint_vis}, this transformation converts abstract protocol structures into rich, high-contrast visual textures, enabling the model to perceive traffic patterns as distinct image features.

\subsection{Traffic Editing with Generative Editor}
\label{sec:method_diffusion}
The Generative Editor executes the actual traffic transformation. It proceeds in three sequential stages: 1) a dual-conditioned backbone integrates paired flow-instruction inputs; 2) a supervised fine-tuning strategy injects domain-specific evasion knowledge into the model; and 3) a controllable inference mechanism regulates the trade-off between evasion efficacy and protocol integrity.

\noindent\textbf{Dual-Conditioned Diffusion Backbone}
To instantiate the instruction-guided editing paradigm defined in \S\ref{sec:method_paradigm}, we employ a conditional latent diffusion architecture capable of dual-input processing. Unlike standard text-to-image models that generate content solely from noise, our task requires an architecture that accepts both an initial image (the censored flow) and a semantic instruction (the evasion prompt). We therefore adopt a \textit{text-conditioned image editing} framework~\cite{Brooks2023-xa} as our generative backbone.

We specifically instantiate this framework using a high-capacity foundation model, \textit{InstructPix2Pix-sdxl}~\cite{sdxl_instructpix2pix_hf}, to handle the intricate dependencies of network protocols. The rationale for this design choice is the strict structural correlation inherent in bit streams: a modification in one header field often necessitates consistent changes in distant fields (e.g., a TCP Data Offset change implies an adjustment in TCP Options). A foundation model with a massive parameter space and dual-encoder design provides the requisite capacity to capture these long-range bit-level correlations. Through iterative denoising, the model predicts the noise residual needed to shift the flow's statistical texture toward a benign pattern, modifying attributes such as TCP options and flag distributions while preserving the protocol structure of the original flow via the image conditioning mechanism.

\noindent\textbf{Supervised Fine-tuning on Flow Image Pairs}
However, such backbones lack domain-specific knowledge for censorship evasion. To bridge this gap, we fine-tune the U-Net while keeping the VAE and text encoders frozen, using a supervised dataset of flow triplets. Each training sample consists of an \textit{Original Flow Image}, a corresponding \textit{Target Evasive Flow Image}, and a \textit{User Instruction}. The target images serve as the ground truth; they are synthesized by processing the original flows through established evasion frameworks (\textit{Geneva}~\cite{geneva2019} and \textit{UPGen} ~\cite{upgen}). The text instruction acts as a semantic condition. This process teaches the model to map high-level evasion intent to statistical header modifications that align with the benign traffic manifold, rather than replicating the specific strategy logic from the 
training data, enabling generalization to unseen protocols beyond what the training strategies cover.

\noindent\textbf{Inference Pipeline and Controllable Generation}
Once the model is fine-tuned, the final stage is the inference process, where we generate evasive flows. The system operates on two user-provided inputs: a raw censored flow encoded as a flow image and a text instruction specifying the target censor, requiring no manual fitness functions, protocol specifications, or per-run training. To precisely control the trade-off between evasiveness and connectivity, we employ a \textit{Dual Classifier-Free Guidance (CFG)} strategy~\cite{ho2022classifier}. This mechanism allows us to independently tune two weighting factors during generation. The text guidance scale dictates how aggressively the model modifies the traffic features to adhere to the evasion command, thereby enhancing evasion success. Conversely, the image guidance scale constrains the generation to preserve the structural skeleton of the original input flow, ensuring protocol validity. By balancing these two scales, \emph{FlowPaint} injects the necessary evasion patterns through in-place feature modification rather than dummy packet injection to bypass detection, preserving sufficient original characteristics and minimizing bandwidth overhead.

\begin{figure}[t]
    \centering
    \includegraphics[width=0.8\linewidth]{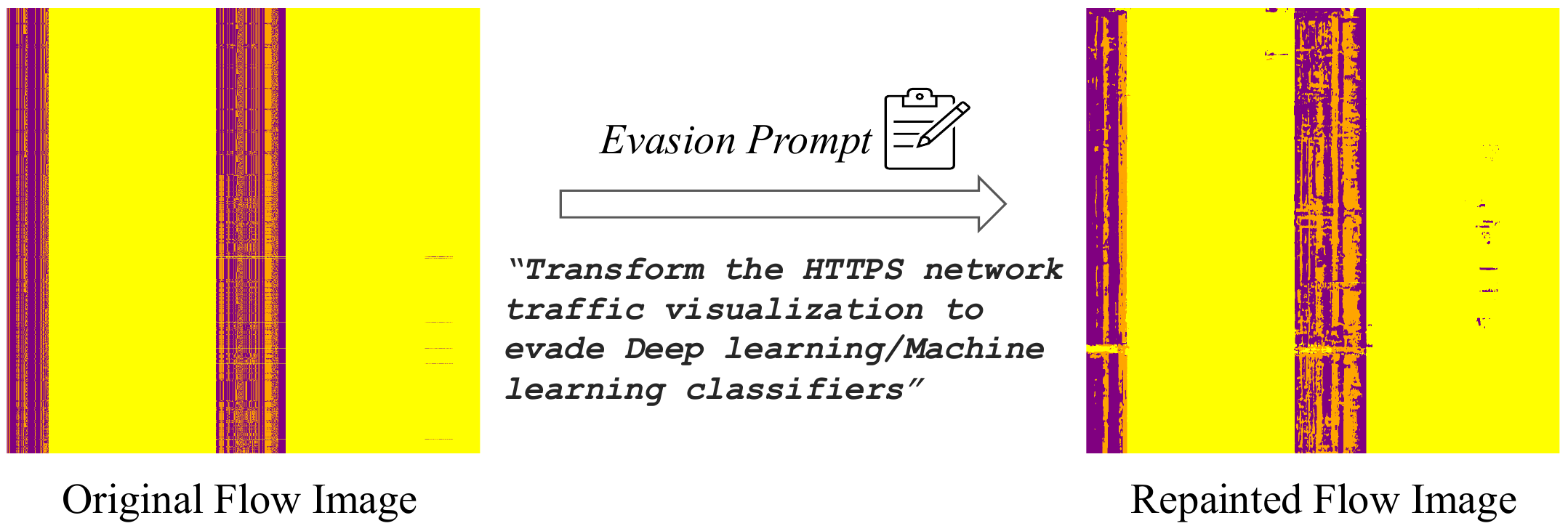} 
     \caption{\textbf{Visualization of Instruction-Guided Traffic Editing.} The inference process: (1) Input \textit{Original Flow} (left); (2) Textual \textit{Evasion Prompt} (middle); and (3) The \textit{Repainted Flow} (right) generated by Generative Editor.}
    \label{fig:repaint_vis}
\end{figure}

Figure~\ref{fig:repaint_vis} visualizes the inference process of our fine-tuned model. On the left, the \textit{Original Flow Image} captures the raw, censored flow signature by traffic encoder, where Orange represents active bits ($1$), Purple denotes unset flags ($0$), and Yellow indicates padding ($-1$). The center panel displays the \textit{Evasion Prompt}, providing the high-level instruction that guides the editing. The \textit{Repainted Flow Image} on the right demonstrates the model's generative output: specific pixel regions have been synthesized to match the manifold of benign traffic.

\subsection{Reconstruction with Protocol Decoder}
\label{sec:method_postprocess}
The output of the diffusion model is a "denoised" image where pixel values represent continuous probabilities rather than discrete bits. This probabilistic nature poses a significant challenge: \textit{hallucination}. For instance, the model might produce ambiguous fractional values for strictly binary TCP flags, or generate checksums that mathematically contradict the accompanying payload. Such artifacts would render the traffic invalid. To resolve this, we employ a hierarchical reconstruction pipeline that systematically projects the probabilistic output back onto the manifold of valid protocols, proceeding from visual decoding to structural repair and finally to stateful restoration. 

\noindent\textbf{Inverse Visual Decoding}
The reconstruction pipeline begins by addressing the domain mismatch between the continuous generative output and discrete network protocols. We invert the ternary color encoding by applying nearest-neighbor classification to the generated RGB pixels, converting orange, purple, and yellow back to active bits ($1$), unset bits ($0$), and placeholders ($-1$), recovering the discrete \emph{nPrint} matrix required for subsequent protocol-level repair.

\noindent\textbf{Intra-packet Repair}
While the previous step recovers the discrete bit values, raw binary data does not guarantee syntactic correctness at the protocol level. 
Consequently, we scan the recovered matrix to enforce structural validity at the individual packet level. 
We apply deterministic fixes including header standardization, which enforces fixed values for protocol versions and ensures strictly non-zero TTLs to prevent routing loops, and field consistency, which recalculates header lengths and total length fields to match the actual generated packet size. Checksums are recalculated automatically during final PCAP reconstruction. This ensures that each row in the matrix represents a syntactically valid network packet.

\noindent\textbf{Inter-packet Consistency Restoration}
Even with syntactically correct individual packets, the flow may still fail due to violations of temporal logic, as network flows are stateful sequences rather than independent packets. The diffusion model, having edited the flow as a static spatial image, may inadvertently disrupt these time-dependent relationships. Therefore, we implement a stateful reconstruction logic to restore consistency across the entire connection timeline. 
First, we use a Markov chain trained on reference traffic to sample a direction sequence for the reconstructed flow, assigning src/dst IP pairs accordingly to each packet based on its inferred client-to-server or server-to-client direction. Second, we perform TCP state machine emulation by recalculating sequence numbers to guarantee valid state transitions, such as ensuring sequence numbers mathematically follow the SYN $\rightarrow$ SYN-ACK $\rightarrow$ ACK progression.

\noindent\textbf{Flow Reassembly}
With the header sequences fully sanitized both spatially and temporally, \textit{Protocol Decoder} executes the final reconstitution of the flow. We convert the repaired \emph{nPrint} matrix back into standard PCAP format. Crucially, this is the stage where we reintegrate the original application payloads, which were explicitly preserved and left untouched during the header-focused editing process. This final reassembly ensures that while the traffic metadata and statistical footprint are modified to evade censorship, the underlying data transmission remains intact, guaranteeing the reliability of the established communication channel.

\section{Implementation}
\label{sec:implementation}
We implemented \emph{FlowPaint} in Python (v3.11), totaling approximately 4,500 lines of code (LoC). This codebase comprises \textit{Traffic Encoder} (\S\ref{sec:method_preprocess}, $\approx$1,200 LoC), \textit{Generative Editor} (\S\ref{sec:method_diffusion}, $\approx$1,500 LoC), and \textit{Protocol Decoder} (\S\ref{sec:method_postprocess}, $\approx$1,800 LoC). \textbf{We plan to open-source \emph{FlowPaint} upon acceptance.}

\noindent\textbf{Training Dataset and Flow Encoding.}
We constructed our training dataset by pairing benign traffic from \textit{Stratosphere IPS}~\cite{stratosphere} with its evasive counterparts. We selected this dataset for its rigorous collection methodology and high fidelity, offering a diverse representation of real-world legitimate traffic. To ensure a comprehensive representation of evasion logic, we synthesized the target evasive flows using two state-of-the-art tools: \textit{Geneva}~\cite{geneva2019}, specifically the strategy \path{[TCP:flags:PA]-duplicate(tamper{TCP:dataofs:replace:10}(tamper{TCP:chksum:corrupt},),)-|}  with a $98\%$ bypass rate against the GFW; and \textit{UPGen}~\cite{upgen}, employing six public Protocol Specification Files (PSFs) to introduce varied structural obfuscation patterns. The final consolidated dataset consists of 6,712 paired flow samples. For traffic encoding, raw PCAP traces are segmented into blocks of 1024 packets, and the \emph{nPrint} encoding produces a feature matrix where the width corresponds to the 1088 header bits.

\noindent\textbf{Training Configuration.} 
We fine-tuned \path{diffusers/sdxl-instructpix2pix-768}~\cite{sdxl_instructpix2pix_hf} on an Intel Core Ultra 9 285K and NVIDIA RTX 5090 (32GB). We froze the VAE ($\approx$83M) and Text Encoders ($\approx$817M), training only the U-Net ($\approx$2.6B) for 10 epochs. We used a learning rate of $1\times 10^{-5}$ (constant schedule, 500 warmup), global batch size 4, and gradient accumulation 4. Optimization employed 8-bit Adam, BF16 precision, and 0.1 conditioning dropout.

\noindent\textbf{Inference Parameters.}
The inference engine utilizes a DDIM sampler with 10 inference steps and empirically optimized guidance scales. Specifically, we set the Text Guidance Scale to 7.5 to ensure high adherence to the evasion prompt, and the Image Guidance Scale to 1.5 to maintain moderate adherence to the original structure. 

\noindent\textbf{Proxy Deployment.} \emph{FlowPaint} operates as a proxy shaping layer: it intercepts outgoing packets via Netfilter, buffers them into 1,024-packet blocks, runs the diffusion pipeline, and re-injects repainted packets via tcpreplay, while maintaining OS-level TCP state independently. Flows shorter than 1,024 packets are zero-padded and trimmed post-decoding.

\section{Evaluation}
\label{sec:evaluation}
We evaluate the evasion robustness of \emph{FlowPaint} against two censorship mechanisms. First, we demonstrate resilience to \textbf{learning-based traffic analysis} using machine learning and deep learning classifiers that simulate statistical adversaries. Second, we verify evasion efficacy against \textbf{rule-based filtering} via deterministic middleboxes modeled after the operational principles of the GFW, representing industrial-grade inspection systems. 

\subsection{Resilience to Learning-based Traffic Analysis}
We evaluate resilience against advanced learning-based traffic analysis, which scrutinize flow attributes to identify evasion attempts via complex, non-linear statistical boundaries. Once a flow is classified as a prohibited protocol with high confidence, the censor triggers a blocking action.

\vspace{-2ex}
\subsubsection{Experimental Setup}

\noindent\textbf{Classifiers.}
We adopt the identical classifier suite from \textit{UPGen}~\cite{upgen} to ensure direct comparability, using \emph{nPrint}~\cite{nprintml} as the standardized input representation: \emph{Decision Tree}, \emph{Random Forest}~\cite{RandomForest}, \emph{nPrintML}~\cite{nprintml} (an AutoML framework), and \emph{Deep Fingerprinting}~\cite{deepfingerprinting}, a 1D Convolutional Neural Network (CNN) originally optimized for website fingerprinting. \emph{Decision Tree} and \emph{Random Forest} represent traditional statistical methods, \emph{Deep Fingerprinting} represents deep learning-based analysis, and \emph{nPrintML} represents AutoML-based ensemble classification, together covering the main families of learning-based traffic analysis approaches.

\noindent\textbf{Baselines.}
We compare \emph{FlowPaint} against two baselines: \textbf{\textit{UPGen}}~\cite{upgen} and \textbf{\textit{Obfs4}}~\cite{obfs4}. 
\textit{UPGen} automatically constructs unknown encrypted protocols to structurally resemble benign traffic, representing the state-of-the-art in resisting learning-based traffic analysis. \textit{Obfs4} serves as the widely deployed industry standard for fully encrypted protocols (FEPs) used by the Tor network. It relies on rigorous cryptographic randomization to mask all protocol metadata.

\noindent\textbf{Out-of-Distribution (OOD) Evaluation.}
To simulate the realistic actions of censors, we adopted an out-of-distribution evaluation method.
We trained classifiers on a dataset containing the specific evasion protocol (\emph{FlowPaint}, \textit{UPGen}, Obfs4) and a subset of benign encrypted protocols, while designating one benign protocol as out-of-distribution. This out-of-distribution protocol is explicitly excluded from the training dataset but included in the test dataset to verify if the classifier incorrectly flags unknown benign traffic as censored.

Specifically, the benign encrypted protocols selected are among the most widely used standards in the current Internet landscape:
\textbf{CurveZMQ}~\cite{curvezmq}: the underlying security protocol for ZeroMQ.
\textbf{secio}~\cite{secio}: a secure transport protocol used by IPFS.
\textbf{SSH}~\cite{ssh}: the secure shell protocol (RFC 4253).
\textbf{TLS}~\cite{tls12, tls13}: transport layer security (versions 1.2 and 1.3).

\noindent\textbf{Dataset.}
To ensure a fair comparison, we constructed parallel datasets for \emph{FlowPaint}, \textit{UPGen}, and \textit{Obfs4} across all four out-of-distribution evaluations.

\textit{Training dataset:} 800 positive and 2,400 negative samples (800 per in-distribution benign protocol).

\textit{Testing dataset:} 200 positive and 800 negative samples (200 per benign protocol).

\textit{Positive Samples.} Positive samples differ across methods:

\emph{FlowPaint}: Generated using the edit prompt: \textit{``Transform the HTTPS network traffic visualization to evade Deep learning/Machine learning classifiers.''} 
This prompt encodes the adversary type rather than the wire protocol, and is applied uniformly across all four OOD protocols to evaluate cross-protocol generalization.

\textit{UPGen}: Sourced from the official \textit{UPGen} dataset~\cite{upgen}, this subset comprises traffic processed by thousands of PSFs, representing a wide spectrum of generative encrypted protocols.

\textit{Obfs4}: Consists of real Tor bridge traffic utilizing the obfs4 obfuscation protocol. These traces were also acquired from the \textit{UPGen} public dataset~\cite{upgen}.

\textit{Negative Samples.} Negative samples were drawn uniformly from four benign encrypted protocols.

\begin{table}[t]
\centering
\renewcommand{\arraystretch}{1.0}
\setlength{\tabcolsep}{4pt}
\resizebox{0.58\columnwidth}{!}{
    \begin{tabular}{ll ccc}
        \toprule
        \multirow{2}{*}{\makecell{\textbf{OOD} \\ \textbf{Protocol}}} & 
        \multirow{2}{*}{\textbf{Classifier}} & 
        \multicolumn{3}{c}{\textbf{TPR (Fixed FPR $\le$ 0.1\%)}} \\
        \cmidrule(lr){3-5}
         & & \textbf{\emph{FlowP}} & \textbf{\textit{UPGen}} & 
         \textbf{\textit{Obfs4}} \\
        \midrule
        \multirow{4}{*}{\textit{CurveZMQ}} 
          & Decision Tree       & \textbf{0.00} & 0.99 & 1.00 \\
          & Random Forest       & \textbf{0.99} & 1.00 & 1.00 \\  
          & nPrintML            & \textbf{0.00} & 1.00 & 1.00 \\
          & Deep Fingerpr.      & \textbf{0.92} & 1.00 & 1.00 \\ 
        \midrule
        \multirow{4}{*}{\textit{secio}} 
          & Decision Tree       & \textbf{0.00} & 0.00 & 1.00 \\
          & Random Forest       & \textbf{0.99} & 1.00 & 1.00 \\
          & nPrintML            & \textbf{0.62} & 1.00 & 1.00 \\
          & Deep Fingerpr.      & \textbf{0.86} & 0.87 & 1.00 \\
        \midrule
        \multirow{4}{*}{\textit{SSH}} 
          & Decision Tree       & \textbf{0.00} & 0.00 & 1.00 \\
          & Random Forest       & \textbf{0.95} & 1.00 & 1.00 \\
          & nPrintML            & \textbf{0.99} & 1.00 & 1.00 \\
          & Deep Fingerpr.      & \textbf{0.80} & 0.92 & 1.00 \\
        \midrule
        \multirow{4}{*}{\textit{TLS}} 
          & Decision Tree       & \textbf{0.00} & 0.36 & 0.00 \\
          & Random Forest       & \textbf{0.94} & 1.00 & 1.00 \\
          & nPrintML            & \textbf{0.00} & 1.00 & 1.00 \\
          & Deep Fingerpr.      & \textbf{0.85} & 0.92 & 0.94 \\
        \bottomrule
    \end{tabular}
}
\caption{\textbf{TPR Comparison.} Risk-averse classifiers, across four OOD 
protocols.}
\label{tab:operational_evasion}
\end{table}

\noindent\textbf{Performance Metrics}
We employ standard binary classification metrics: \textbf{True Positive Rate (TPR)}, defined as the fraction of evasion flows correctly blocked by the censor, and \textbf{False Positive Rate (FPR)}, the fraction of benign flows incorrectly classified as evasion traffic. 

A successful evasion method is characterized by \textbf{minimizing the TPR while maximizing the FPR}. A relatively low TPR indicates that a significant amount of evasion traffic successfully hides its signatures, slipping through the classifier undetected. Simultaneously, a high FPR is equally critical as it exploits the \textit{Censor's Dilemma}. Real-world censors operate under strict constraints regarding network utility; even a marginally elevated FPR translates to the disruption of significant amounts of legitimate communications. This collateral damage is untenable for state-level censors, who typically prioritize availability for economic stability~\cite{ensafi2015examining}. 

\subsubsection{Evaluation Methodology and Results}
\label{sec:evaluation_scenarios}
We conduct our evaluation under two scenarios.
The first scenario simulates a realistic production environment constrained by strict false positive limits, while the second assesses the fundamental distinguishability of flow features against detectors operating at maximum sensitivity.

\noindent\textbf{Risk-Averse Classifiers.}
This setting simulates a realistic deployment where classifiers must operate under strict constraints to maintain network utility. We enforce a strict budget of \textbf{FPR $\le$ 0.1\%} across all classifiers. 
Instead of utilizing default decision boundaries, we dynamically calibrate the classification threshold for each classifier using only the benign traffic distribution, requiring no ground-truth evasion labels and ensuring that 99.9\% of legitimate flows are correctly preserved.

Table~\ref{tab:operational_evasion} presents comparative detection rates (TPR) of \emph{FlowPaint} against baselines. Under strict constraints, \emph{FlowPaint} demonstrates exceptional efficacy. Notably, it achieves perfect evasion \textbf{(TPR of 0.00)} against \textit{Decision Tree} and \emph{nPrintML} across multiple OOD protocols. This stands in stark contrast to \textit{Obfs4} and \textit{UPGen}, which remain almost entirely exposed (TPR $\approx$ 1.00) in similar settings, confirming that our approach successfully eliminates the explicit structural fingerprints and byte-level statistics that these classifiers target.

This advantage persists against sophisticated deep learning and ensemble models, where \emph{FlowPaint} outperforms baselines by eroding detection confidence. While \textit{Obfs4} remains fully detectable by \emph{Deep Fingerprinting} (TPR of 1.00), \emph{FlowPaint} successfully suppresses detection rates---\allowbreak reducing TPR to \textbf{0.80} on SSH. 
Even against \textit{Random Forest}, which exhibits extreme resilience, \emph{FlowPaint} emerges as the sole strategy capable of breaking the ceiling of absolute certainty. Whereas baselines yield a flat 1.00 TPR, \emph{FlowPaint} forces the detector to miss a measurable fraction of flows (lowering TPR to \textbf{0.94--0.95}), demonstrating its unique ability to perturb traffic features beyond the strict boundaries of benign training data.

\begin{figure*}[t!]
    \centering
    \includegraphics[width=\textwidth]{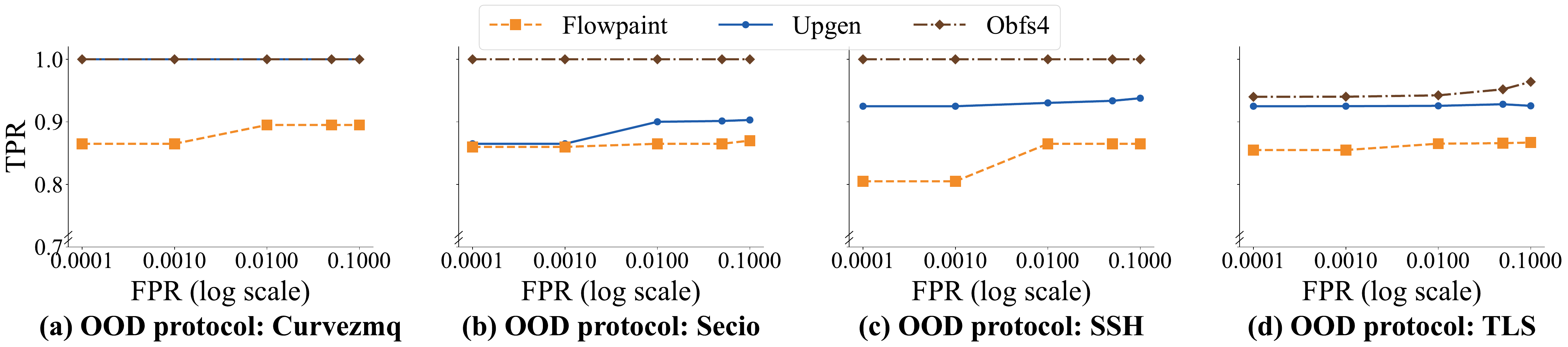} 
     \vspace{-5ex}
    \caption{\textbf{Deep Fingerprinting Sensitivity.} TPR of four OOD protocols at varied FPR thresholds ($10^{-4}$, $10^{-3}$, $0.05$, $0.1$).}
    \label{fig:df_roc}
\end{figure*}

To further scrutinize evasion performance under varying strictness levels, we conduct a granular ROC analysis on \emph{Deep Fingerprinting}. We select \emph{Deep Fingerprinting} as the representative model due to its proven efficacy and widespread adoption in learning-based traffic analysis~\cite{deepfingerprinting}. Figure~\ref{fig:df_roc} visualizes TPR against a logarithmic scale of FPR, explicitly examining performance at critical operational thresholds (e.g., $10^{-4}$, $10^{-3}$, $0.05$, $0.1$). \emph{FlowPaint} consistently exhibits its superior evasion capability, represented by the lowest TPR curves compared to \textit{UPGen} and \textit{Obfs4}. This indicates that \emph{FlowPaint} is the most robust strategy against deep learning-based classification, forcing the classifier to accept operationally prohibitive FPRs in order to achieve meaningful detection.

\noindent\textbf{High-Sensitivity Classifiers.}
We also analyze classifiers using default decision boundaries. This simulates maximum sensitivity, prioritizing evasion detection over minimizing false positives, to assess statistical indistinguishability under maximum discriminatory power.

Table~\ref{tab:comprehensive_resilience} shows the TPR and FPR results for four classifiers across all OOD protocols. The results demonstrate that \emph{FlowPaint} consistently blurs the line between censored and benign flows, creating operational dilemmas for all classifiers. Against ensemble and statistical models, it triggers an FPR of approximately \textbf{25\%} across all OOD protocols for \textit{Random Forest}. While against \emph{nPrintML}, it frequently drives the FPR to \textbf{1.00}. This confirms that their traffic retains distinct artifacts, allowing classifiers to filter them out precisely without the risk of collateral damage or detection uncertainty. In contrast, the baselines fail to create either of these challenges. \textit{UPGen} and \textit{Obfs4} typically exhibit a "clean separation" pattern—consistently maintaining a high TPR (often 1.00) while keeping the FPR at 0.00. This resilience extends to \emph{Deep Fingerprinting}, where \emph{FlowPaint} demonstrates superior evasion capability by effectively suppressing detection rates. While \textit{Obfs4} and \textit{UPGen} remain highly vulnerable to this deep learning model (TPR $\ge$ 0.90), \emph{FlowPaint} successfully erodes detection confidence, lowering the TPR to roughly \textbf{0.70--0.80}. 

\subsubsection{Analysis}
\noindent\textbf{The Limitation of Entropy-Based Obfuscation (\textit{Obfs4}) and Procedural Assembly (\textit{UPGen}).}
\textit{Obfs4}'s reliance on randomized payloads creates high-entropy fingerprints that stand out distinctively against the structured, low-entropy headers of benign protocols. As Table~\ref{tab:comprehensive_resilience} confirms, even a Decision Tree achieves TPR $\approx$ 1.0 with FPR $\approx$ 0.0 against Obfs4. \textit{UPGen} improves upon randomization by procedurally assembling protocols to statistically blend in with benign traffic. However, while it captures explicit protocol syntax, procedural rules inevitably leave subtle sequential artifacts or rigid statistical patterns that remain detectable at similar rates.

\noindent\textbf{The Advantage of Manifold Learning (\emph{FlowPaint}).}
In contrast, \emph{FlowPaint} learns the high-dimensional latent manifold of benign traffic, capturing complex, non-linear statistical correlations that place generated flows deep within the benign feature space. 
While \textit{UPGen} replicates the structural \textit{skeleton} of protocols, \emph{FlowPaint} synthesizes the authentic \textit{texture}, the subtle statistical variations that elude even ensemble classifiers. Critically, the elevated TPR of \textit{Random Forest} in Table~\ref{tab:operational_evasion} represents an oracle upper bound: the threshold is calibrated on the exact test-time benign distribution. 
Under realistic deployment conditions, the same RF classifier must accept $\mathbf{\sim}$\textbf{25\%} FPR to maintain high TPR, as confirmed by Table~\ref{tab:comprehensive_resilience}, blocking one in four legitimate flows, a level of collateral damage no state-level censor can sustain~\cite{ensafi2015examining}.

\begin{table}[t]
\centering
\renewcommand{\arraystretch}{1.05}
\setlength{\tabcolsep}{5pt}
\resizebox{0.7\columnwidth}{!}{
    \begin{tabular}{ll cc cc cc}
        \toprule
        \multirow{2}{*}{\makecell{\textbf{OOD} \\ \textbf{Protocol}}} & 
        \multirow{2}{*}{\textbf{Classifier}} & 
        \multicolumn{2}{c}{\textbf{\emph{FlowPaint}}} & 
        \multicolumn{2}{c}{\textbf{\textit{UPGen}}} & 
        \multicolumn{2}{c}{\textbf{\textit{Obfs4}}} \\
        \cmidrule(lr){3-4} \cmidrule(lr){5-6} \cmidrule(lr){7-8}
         & & \textbf{TPR} & \textbf{FPR} & \textbf{TPR} & \textbf{FPR} 
         & \textbf{TPR} & \textbf{FPR} \\
        \midrule
        \multirow{4}{*}{\textit{CurveZMQ}} 
          & Decision Tree      & \textbf{0.99} & \textbf{0.26} & 1.00 & 0.16 & 1.00 & 0.00 \\
          & Random Forest      & \textbf{1.00} & \textbf{0.25} & 1.00 & 0.00 & 1.00 & 0.00 \\
          & nPrintML           & \textbf{1.00} & \textbf{1.00} & 1.00 & 0.00 & 1.00 & 0.00 \\
          & Deep Fingerpr.     & \textbf{0.80} & \textbf{0.00} & 0.99 & 0.00 & 0.88 & 0.00 \\
        \midrule
        \multirow{4}{*}{\textit{secio}} 
          & Decision Tree      & \textbf{1.00} & \textbf{0.05} & 0.99 & 0.00 & 1.00 & 0.00 \\
          & Random Forest      & \textbf{1.00} & \textbf{0.25} & 1.00 & 0.00 & 1.00 & 0.00 \\
          & nPrintML           & \textbf{1.00} & \textbf{1.00} & 1.00 & 0.00 & 1.00 & 0.15 \\ 
          & Deep Fingerpr.     & \textbf{0.70} & \textbf{0.00} & 0.93 & 0.25 & 1.00 & 0.05 \\
        \midrule
        \multirow{4}{*}{\textit{SSH}} 
          & Decision Tree      & \textbf{1.00} & \textbf{0.26} & 1.00 & 0.25 & 1.00 & 0.25 \\
          & Random Forest      & \textbf{1.00} & \textbf{0.25} & 1.00 & 0.00 & 1.00 & 0.00 \\
          & nPrintML           & \textbf{1.00} & \textbf{1.00} & 1.00 & 0.00 & 1.00 & 0.00 \\
          & Deep Fingerpr.     & \textbf{0.71} & \textbf{0.00} & 0.92 & 0.00 & 1.00 & 0.00 \\
        \midrule
        \multirow{4}{*}{\textit{TLS}} 
          & Decision Tree      & \textbf{1.00} & \textbf{0.26} & 1.00 & 0.25 & 1.00 & 0.25 \\
          & Random Forest      & \textbf{1.00} & \textbf{0.25} & 1.00 & 0.25 & 1.00 & 0.25 \\
          & nPrintML           & \textbf{1.00} & \textbf{1.00} & 1.00 & 0.00 & 1.00 & 0.00 \\
          & Deep Fingerpr.     & \textbf{0.71} & \textbf{0.00} & 0.92 & 0.00 & 1.00 & 0.26 \\
        \bottomrule
    \end{tabular}
}
\caption{\textbf{TPR and FPR Comparison.} High-sensitivity classifiers, 
across four OOD protocols.}
\label{tab:comprehensive_resilience}
\end{table}

\subsection{Resilience to Rule-based Filtering}
To assess the evasion robustness, we deployed three censors acting as ``server-side'' middleboxes. They operate on deterministic inspection logic, scrutinizing packet headers and payloads for specific static signatures. Each censor represents a distinct real-world inspection strategy, from stateful keyword detection to selective compliance checking and active replay defense.

\subsubsection{Censor Configurations}
We instantiate three distinct censor configurations representing different inspection strategies and blocking mechanisms.

\textbf{Censor A: Classic Stateful Reassembler.} 
This model emulates the standard operating mode of the GFW as documented in prior studies~\cite{geneva2019, imc17middlebox}. It maintains a Transmission Control Block (Pseudo-TCB) to track connection states and reassemble TCP streams. Aligning with the GFW's tendency to conserve resources and avoid collateral damage, Censor A adopts a ``fail-open'' policy for malformed packets: it validates TCP checksums and data offsets, ignoring invalid packets rather than terminating the connection. It only triggers a connection tear-down (via RST injection) when a reassembled payload strictly matches a forbidden keyword.

\textbf{Censor B: Heuristic Header Filter.} 
Reflecting the GFW's observed behavior of ceasing inspection on unrecognized protocol variants, this model applies selective RFC compliance checking. It stops tracking flows whose TCP data offset exceeds standard bounds ($\ge$ 9), treating them as unsupported protocol extensions and stopping keyword detection. While such flows pass uninspected, this effectively excludes evasion strategies that rely on large header padding from keyword detection, as those flows are no longer tracked.

\textbf{Censor C: Anti-Replay Validator.} 
This configuration mimics the GFW's sophisticated active countermeasures against probing and replay attacks~\cite{gfw-replay-attack}. It caches $(SEQ,$ \allowbreak $Payload\_Hash)$ pairs to detect TCP segment replays and duplicate payloads. If such artifacts are detected, mirroring the GFW's response to detected Tor bridges or Shadowsocks probes, the censor actively flags the flow and injects bidirectional RSTs, closing the evasion window. 

Table~\ref{tab:censor_configs} provides a comparative summary of three configurations.
To ensure rigorous comparison, Censor A is directly adopted from the open-source mock censor implemented by \textit{Geneva}~\cite{geneva2019}, whereas Censors B and C are custom-constructed to simulate more aggressive and sophisticated filtering behaviors observed in recent measurements.  

\subsubsection{Evasion Effectiveness}
We evaluated the efficacy of both \emph{FlowPaint} and \textit{Geneva} against the three censor configurations by initiating HTTPS requests containing the sensitive keyword 'ultrasurf'.
To equip \emph{FlowPaint} with evasion capabilities, we fine-tuned our model on a paired dataset constructed from benign HTTPS traffic (sourced from Stratosphere IPS~\cite{stratosphere}) and their obfuscated counterparts generated via \textit{Geneva}'s strategies against the GFW. During the evaluation, we guided the model's inference process using the specific text prompt: \textit{`Transform the HTTPS network traffic visualization to bypass the GFW'}, selected to match the rule-based censorship modality.

For \textit{Geneva}, we utilized a widely successful open-source strategy against the GFW: \path{[TCP:flags:PA]-duplicate(tamper{TCP:dataofs:replace:10}(tamper{TCP:chksum:corrupt},),)-|}. This strategy attempts to confuse the censor by sending a "decoy" packet with a corrupt checksum and modified data offset before sending the real packet. \textit{UPGen} and \textit{Obfs4} are excluded as they target ML classifiers rather than rule-based keyword filters.

The results are summarized in Table~\ref{tab:evasion_results}. Both strategies successfully evaded Censor A and B. However, a divergence in efficacy was observed against Censor C. 
\begin{table*}[t]
    \centering
    \small 
    \renewcommand{\arraystretch}{1.35} 
    \setlength{\tabcolsep}{6pt}     
    
    \resizebox{\textwidth}{!}{
    \begin{tabular}{lllll}
        \toprule
        \textbf{Model} & \textbf{Target} & \textbf{Inspection Mechanism (Logic)} & \textbf{Handling Policy} & \textbf{Blocking Trigger} \\
        \midrule
        
        \rowcolor{gray!10} \textbf{Censor A} 
        & Standard
        & Stateful TCP reassembly (Pseudo-TCB)
        & \textbf{Fail-Open} (Permissive)
        & Keyword in payload \\
        
        \textbf{Censor B} 
        & Heuristic
        & Selective RFC compliance check
        & \textbf{Fail-Open} (On Anomalies)
        & Keyword in tracked flows  \\
        
        \rowcolor{gray!10} \textbf{Censor C} 
        & Active
        & Multi-flow Correlation \& Replay Detection
        & \textbf{Active Defense} (RST Injection)
        & Replay patterns \\
        
        \bottomrule
    \end{tabular}
    }
    \caption{\textbf{Censor Configurations.} Comparison of inspection mechanisms and blocking policies.}
    \label{tab:censor_configs}
\end{table*}

\subsubsection{Analysis}
The performance divergence stems from the fundamental architectural difference between \textit{Geneva}'s \emph{Protocol Ambiguity} strategy and \emph{FlowPaint}'s \emph{Traffic Repainting}. 
\textit{Geneva} injects duplicate packets with malformed artifacts (corrupted checksums, anomalous data offsets) to desynchronize the censor's state. Both strategies evade Censor A and B, but through different mechanisms: against Censor A, \emph{FlowPaint} alters TCP options and data offset fields to shift payload boundaries, rendering the keyword invisible to the reassembler without injecting any extra packets; against Censor B, \textit{Geneva} exploits the fail-open behavior by pushing data offset above the threshold, while \emph{FlowPaint} keeps data offset within standard bounds and avoids triggering the heuristic entirely. \textit{Geneva} fails against Censor C because its duplicate injections and malformed artifacts simultaneously trigger the blocking conditions. \emph{FlowPaint} avoids these: the Protocol 
Decoder recalculates checksums to produce valid traffic, data offset remains within standard bounds, and no duplicate packets are injected. In essence, \emph{FlowPaint} succeeds not by exploiting censor implementation bugs, but by producing traffic that is intrinsically indistinguishable from benign flows.

\begin{table}[h]
    \centering
    \small
    \setlength{\tabcolsep}{2.5pt}
    \begin{tabular}{lccc}
        \toprule
        \textbf{Method} & \textbf{\shortstack{Censor A \\ (Stateful)}} & \textbf{\shortstack{Censor B \\ (Header Drop)}} & \textbf{\shortstack{Censor C \\ (Anti-Replay)}} \\
        \midrule
        \rowcolor{gray!10} Geneva~\cite{geneva2019} & \cmark & \cmark & \xmark \\
        \textbf{FlowPaint} & \textbf{\cmark} & \textbf{\cmark} & \textbf{\cmark} \\
        \bottomrule
    \end{tabular}
    \caption{Evasion Efficacy against Three Censors. (\cmark: bypass, \xmark: blocked).}
    \label{tab:evasion_results}
    \vspace{-3ex}
\end{table}

\vspace{0.5ex}
\section{Ablation Study}
\label{sec:ablation}

\noindent\textbf{Importance of image representation and pre-trained knowledge.}
To understand whether the image-based design of \emph{FlowPaint} is essential, we evaluate three ablated variants sharing the complete system's training data, schedule, and Protocol Decoder under the same setting as Table~\ref{tab:operational_evasion}: \textit{RandVAE} (random VAE, pre-trained U-Net), \textit{RandAll} (both randomly initialized), and \textit{NoImage}, which discards the image pathway and processes the flattened 1.1M-dimensional bit vector with a 4-layer MLP. All three variants yield \textbf{TPR of 1.00} across all OOD protocols and classifiers, indicating complete evasion failure despite identical training data. The consistent failure of \textit{RandVAE} 
and \textit{RandAll} confirms that pre-trained visual knowledge is not merely helpful but necessary: the image abstraction alone, without transferred weights, is insufficient. The failure of \textit{NoImage} further shows that the two-dimensional layout itself is essential: it preserves the temporal structure of the flow, allows the convolutional backbone to capture multi-scale patterns across packets, and provides a latent bottleneck that regularizes learning over a high-dimensional space with few examples. All of these are lost when a dense MLP treats each bit independently.
 
\noindent\textbf{Contribution of Generative Editor versus Protocol Decoder.} To quantify what the Generative Editor contributes beyond the deterministic repairs of the Protocol Decoder, we ablate from two directions. Removing the Generative Editor, both \textit{Decoder-only} (the original censored flow passed through the Protocol Decoder) and \textit{Fuzzer+Decoder} (random bit flips on editable fields at 10\%, 30\%, and 50\% before decoding) yield a \textbf{TPR of 1.00} across all OOD protocols and classifiers, evaluated under the same setting as Table~\ref{tab:operational_evasion}. Neither deterministic repair nor random perturbation, even flipping half of all editable bits, produces any evasion. 

Removing the Protocol Decoder instead, the \textit{Diffusion-only} output passes basic header checks (verifying that protocol version, header lengths, data offsets, and TTL values are within legal ranges) for \textbf{99.69\%} of packets but converts to valid PCAP for only \textbf{0.00\%}, owing to deeper inconsistencies (checksums, sequence numbers, and length fields) the decoder must repair. Both components are thus necessary: the Generative Editor for evasion, the Protocol Decoder for protocol validity. Of the 1{,}088 bits modified by the Generative Editor, \textbf{75.8\%} fall in protocol-critical fields subsequently repaired by the Protocol Decoder, while the remaining \textbf{24.2\%} constitute effective semantic modifications in statistical-fingerprint fields such as TCP window size and IPv4 identification, as summarized in Table~\ref{tab:bit_labor}. Notably, this \textbf{24.2\%} of effective modifications is less than half the fuzzer's 50\% random modification rate, yet achieves \textbf{perfect evasion (TPR = 0.00)} across multiple OOD protocols and classifiers against the fuzzer's 0\%, showing that evasion is driven by \emph{which} bits change, not how many. The Generative Editor learns precise, semantically valid modifications that random perturbation cannot replicate.
 
\noindent\textbf{Interpreting the evasion mechanism.}
To verify that \emph{FlowPaint} modifies semantically meaningful fields rather than arbitrary bits, we examine which fields it modifies using Random Forest feature importance aggregated~\cite{RandomForest} across the four OOD protocols under the same setting as 
Table~\ref{tab:operational_evasion}. Table~\ref{tab:xai_overlap} shows that all three top fields \emph{FlowPaint} modifies are among the fields the censor classifier 
relies on most, confirming targeted rather than arbitrary editing. 
For all three fields, \emph{FlowPaint} shifts the distribution toward 
benign traffic. We measure improvement as the percentage reduction in 
each field's mean deviation from the benign baseline; the three fields 
achieve an average improvement of \textbf{76.2\%}, confirming that the 
model makes semantically directed edits rather than arbitrary 
perturbations.

\begin{table}[t]
\centering
\renewcommand{\arraystretch}{1.25}
\begin{minipage}[t]{0.56\columnwidth}
    \centering
    \setlength{\tabcolsep}{4pt}
    \small
    \begin{tabular}{@{}>{\centering\arraybackslash}p{0.20\linewidth} >{\centering\arraybackslash}p{0.46\linewidth} r@{}}
        \toprule
        \textbf{Handled by} & \textbf{Key fields} & \textbf{\% Edits} \\
        \midrule
        \rowcolor{gray!10} \textbf{Protocol Decoder}
        & seq, ack, ports; lengths, offsets, version, fragmentation; checksums
        & \textbf{75.8\%} \\
        \textbf{Generative Editor}
        & \texttt{ipv4\_id}; \texttt{tcp\_wsize}; \texttt{tcp\_urp}; \texttt{ipv4\_tos}
        & \textbf{24.2\%} \\
        \bottomrule
    \end{tabular}
    \captionof{table}{\textbf{Generative Editor and Protocol Decoder Contributions.} Allocation of the 1{,}088 bits between the two components.}
    \label{tab:bit_labor}
\end{minipage}
\hfill
\begin{minipage}[t]{0.42\columnwidth}
    \centering
    \renewcommand{\arraystretch}{1.25}
    \small
    \begin{tabularx}{0.7\linewidth}{@{}Xr@{}}   
        \toprule
        \textbf{Field} & \textbf{Improv.} \\
        \midrule
        \rowcolor{gray!10} \texttt{ipv4\_id}   & \textbf{58.4\%} \\
        \rowcolor{gray!10} \texttt{ipv4\_dscp} & \textbf{88.2\%} \\
        \texttt{ipv4\_ttl}  & \textbf{81.9\%} \\
        \bottomrule
    \end{tabularx}
    \captionof{table}{\textbf{Modification Effectiveness} of \emph{FlowPaint} on the censor-relied fields that move toward the benign distribution.}
    \label{tab:xai_overlap}
\end{minipage}
\end{table}

\section{Discussion}
\label{sec:discussion}
\noindent\textbf{The Necessity of \emph{FlowPaint}.}
Could similar evasion capabilities be achieved with simpler models? One alternative would be refining evolutionary algorithms like \textit{Geneva}~\cite{geneva2019}. While effectively exploit specific censor bugs (e.g., TCP state desynchronization), they often produce fragmented "adversarial patches" that lack semantic coherence and fail once bugs are patched. Contemporary deep learning censors analyze holistic traffic patterns rather than isolated fields. To defeat them, an evasion tool must reconstruct complex inter-packet dependencies to project the flow onto a benign manifold. \emph{FlowPaint} employs a diffusion-based backbone that provides the essential capacity for this global reshaping—a level of statistical fidelity that simpler heuristics or evolutionary search cannot theoretically achieve. 

\noindent\textbf{Censor Countermeasures and Co-evolution.}
Censors might react to \emph{FlowPaint} in two ways. First, they may detect generative fingerprints. However, since \emph{FlowPaint} matches benign distributions, detection would require the censor to narrow its acceptance window significantly, increasing unacceptable damage to legitimate traffic. Second, censors might inject perturbations to disrupt generation, sparking a co-evolution with evasion and censorship systems iteratively competing. As censorship strategies evolve, \emph{FlowPaint} can adapt them by fine-tuning on newly generated bypass-verified traffic pairs, handling novel threats through data updates rather than complex re-engineering.  We note that the learned manifold reflects our training distribution; generalization to different network environments requires future cross-dataset validation.

\noindent\textbf{The Cost of Strategic Adaptability.}
\emph{FlowPaint}'s primary trade-off is computational overhead: unlike lightweight systems such as \textit{Obfs4}, our generative diffusion model reduces goodput and introduces an initial inference latency, which we characterize in detail in Appendix~\ref{appendix:performance}. We argue this is a worthwhile trade-off, as we prioritize evasion over raw efficiency; in censorship scenarios where the alternative is total blockage, a setup delay is a negligible price for access. Crucially, this overhead shifts the bottleneck from network conditions to compute resources, turning it into a manageable hardware constraint rather than an algorithmic flaw that hardware acceleration will naturally diminish. Characterizing energy consumption, GPU occupancy, and performance across commodity hardware profiles remains an important direction for future work.

\noindent\textbf{Ethics Considerations}
This research does not involve human subjects or personally identifiable information and therefore does not require IRB review. To prioritize network safety and avoid harm to third parties, we do \emph{not} conduct active measurements against live national firewalls; instead, we evaluate \emph{FlowPaint} in a controlled environment using local filters calibrated from publicly available, openly licensed datasets that contain
no user-identifying information. While such simulations cannot fully capture the complexity of nation-state censors, they validate \emph{FlowPaint}'s core effectiveness without exposing any real user or network to risk. 
We emphasize that \emph{FlowPaint} is intended as a research contribution to advance the understanding of network security and censorship resistance, rather than as an operational tool for attacking real-world censors.

\section{Related Work} \label{sec:related_work}
\noindent\textbf{Censorship Paradigms} Censorship manifests in three primary paradigms. First, rule-based filtering uses DPI and static signatures for deterministic blocking~\cite{Dyer2013-up, Dixon2016-mj}. Second, obfuscation-aware inspection uses entropy and n-gram analysis of encrypted payloads to distinguish generic encryption from manual obfuscation~\cite{Wang2015-ea}. The third paradigm utilizes machine-learning and deep-learning to identify flow-level statistical features, such as packet timing, sizing distributions~\cite{Gong2020-hd}. 
\emph{FlowPaint} overcomes the first and third paradigms by leveraging the semantic depth of diffusion models to ``repaint'' traffic so that it conforms to benign, permitted protocols. 
While \emph{FlowPaint} focuses on modifying packet headers, it does not address censors that analyze the encrypted payload's byte-level randomness.

\noindent\textbf{Traffic Generation.} Generative models in networking have evolved from GAN-based packet synthesis like \textit{PAC-GAN}~\cite{Cheng2019-an} to diffusion-based generation like \textit{NetDiffusion}~\cite{Jiang2024-me}, and language model adaptations like \textit{ET-BERT}~\cite{lin2022etbert} for payload interpretation. The objective of these works is synthesizing new traffic entirely from scratch to address data scarcity. 
Unlike traffic generation, \emph{FlowPaint} performs live traffic editing. This requires selectively transforming specific features of an existing blocked user flow to bypass the censor, while rigorously preserving the protocol semantics and communicative intent.

\noindent\textbf{Autonomous Traffic Reshaping.} Recent studies leverage data-driven intelligence for complex networking tasks, such as \textit{SymTCP}'s~\cite{wang2020symtcp} exploitation of state-machine discrepancies and Xie et al.'s~\cite{xie2026defending} programmable switch framework for defending against traffic analysis. While \emph{FlowPaint} shares the trend of utilizing intelligent agents, it distinguishes itself through generative semantic abstraction. Unlike defensive infrastructure solutions or reinforcement learning (RL) based optimization (which typically seek specific noise vectors), \emph{FlowPaint} leverages the "world knowledge" of diffusion models to reconstruct traffic patterns, prioritizing the high-fidelity protocol semantics over the specific noise vectors sought by gradient-based adversarial methods~\cite{Li2022-vl, Nasr2021-dj}, which additionally require white-box classifier access and fail to generalize across unknown censor architectures.

\section{Conclusion and Future Work}
\label{sec:conclusion}
We propose \emph{FlowPaint}, a framework that first reframes censorship evasion as a semantic image-to-image editing task. By leveraging diffusion-based architecture to "repaint" traffic patterns under natural language guidance, our approach bridges the gap between high-level user intent and low-level packet manipulation. This paradigm moves beyond static heuristics, offering a robust and automated solution that evades both rule-based and learning-based censors. 
\emph{FlowPaint} transforms labor-intensive protocol engineering into a simplified, instruction-driven task, ensuring robust communication against increasingly heterogeneous and hybrid censorship.

Looking forward, we position \emph{FlowPaint} as a complementary shaping layer for the existing privacy ecosystem. Future work will focus on integrating it as a pluggable module atop established tunnels~\cite{obfs4, shadowsocks_web}, augmenting their encryption with dynamic statistical repainting. 
Furthermore, \emph{FlowPaint}'s flexibility enables rapid response to emerging threats; by enabling the community to "patch" vulnerabilities via model weight updates, \emph{FlowPaint} offers a scalable path for rapid adaptation to censorship techniques.

\appendix
\section{Algorithm Goodput and Latency}
\label{appendix:performance}
We evaluate the algorithmic goodput and latency of \emph{FlowPaint} to understand its capabilities relative to distinct evasion strategies and protocol configurations.

\noindent\textbf{Experimental Setup.}
To ensure the credibility of our evaluation, we strictly align our experimental configuration and measurement methodology with those established in prior work~\cite{upgen}. We conduct our experiments using the Shadow discrete-event network simulator~\cite{shadow}, executing unmodified application binaries in a deterministic environment. Our topology consists of a client, a proxy, and a server, each running in a virtual host. Network links are configured with 
sufficient bandwidth to ensure bottlenecks arise from proxy processing rather than link capacity.

\noindent\textbf{Evasion Strategies and Configurations.}
We compare \emph{FlowPaint} against the following transport configurations. \textit{Dummy} is a raw TCP connection without encryption or obfuscation, serving as the theoretical upper bound. \textit{UPGen} Variants are instantiated using the state-of-the-art protocol generator~\cite{upgen} with three distinct PSFs to represent various generative encrypted protocols: \textit{Public Key} represents a heavy configuration involving a bidirectional handshake with 115-byte keys and random padding, simulating a custom encrypted protocol; \textit{TLS Mimic} mimics the TLS 1.2 record structure (ClientHello/ServerHello) to blend in with HTTPS traffic; and \textit{Shadowsocks} serves as a lightweight streaming configuration 
that mimics AEAD frames and initiates data transfer immediately without a handshake.

\noindent\textbf{Goodput.}
We evaluate the steady-state forwarding capacity of each protocol by strictly isolating the proxy's algorithmic overhead from network bandwidth limitations. The client is configured to emulate a bandwidth-saturated user performing a bulk 1 GiB download from the server (paired with minimal 1 KiB upstream control traffic). To ensure the bottleneck resides solely within the evasion logic, we utilize a loopback network topology provisioned with high-capacity bandwidth and a negligible 1 ms Round-Trip Time (RTT). We execute this workload for each protocol configuration and analyze the client's transfer logs to calculate the application-level goodput—defined as the rate of valid payload delivery excluding transport and obfuscation headers.

\begin{figure}[t]
    \centering
    \begin{minipage}[c]{0.38\columnwidth}
        \centering
        \vspace{0.8ex}
        \small
        \renewcommand{\arraystretch}{1.4}
        \setlength{\tabcolsep}{5pt}
        \begin{tabular}{lc}
            \toprule
            \textbf{Protocol} & \textbf{Goodput (Gbps)} \\
            \midrule
            \rowcolor{gray!10} Dummy & 10.85 \\
            \textit{UPGen}: Public Key & 10.80 \\
            \rowcolor{gray!10} \textit{UPGen}: TLS Mimic & 10.75 \\
            \textit{UPGen}: Shadowsocks & 10.72 \\
            \rowcolor{gray!10} \textbf{FlowPaint} & \textbf{3.05} \\
            \bottomrule
        \end{tabular}
        \vspace{0.8ex}
        \captionof{table}{\textbf{Goodput Performance.}}
        \label{tab:throughput}
    \end{minipage}
    \hfill
    \begin{minipage}[c]{0.58\columnwidth}
        \centering
        \includegraphics[width=\linewidth, height=5cm, keepaspectratio=false]{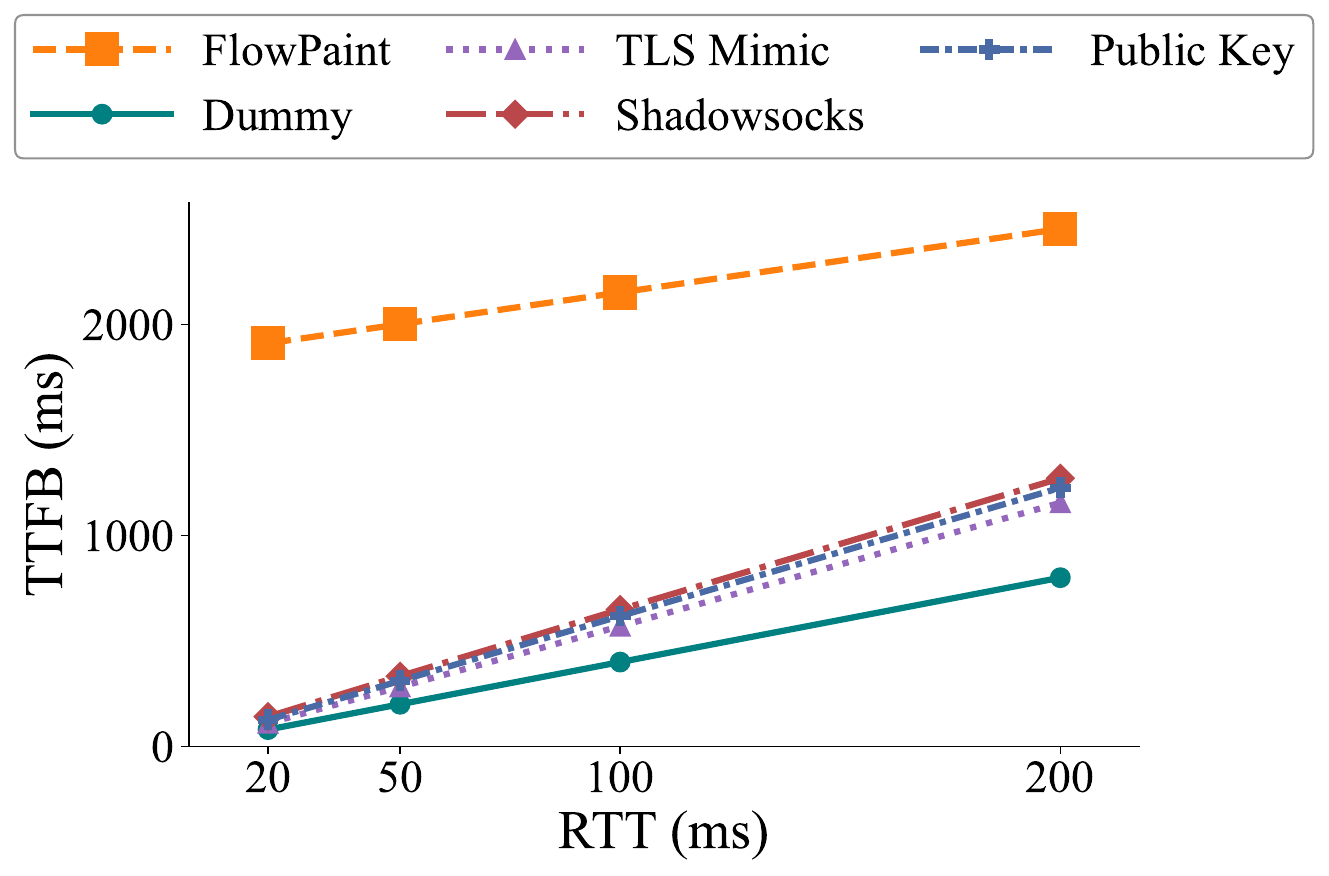}
        \vspace{-4ex}
        \captionof{figure}{\textbf{Latency Performance.} TTFB trends across varying network RTTs.}
        \label{fig:latency}
    \end{minipage}
\end{figure}

Table~\ref{tab:throughput} presents the steady-state goodput results. \textit{Dummy} and \textit{UPGen} saturate the available bandwidth at approximately 10.7--10.85 Gbps, confirming their lightweight processing overhead; however, as shown in Table~\ref{tab:operational_evasion}, both achieve near-zero evasion against ML classifiers (TPR$\approx$1.00), rendering their throughput advantage meaningless in practice. \emph{FlowPaint} achieves a steady-state goodput of \textbf{3.05 Gbps}, a reduction attributed to the computational intensity of the generative 
diffusion model. Despite this overhead, the goodput remains more than sufficient for high-bandwidth applications; with 4K and 8K streaming requiring only 25 Mbps and 100 Mbps respectively~\cite{netflix_speed, youtube_bitrate}, \emph{FlowPaint} exceeds these demands by two orders of magnitude. Crucially, unlike baselines that achieve near-zero evasion, this overhead buys meaningful evasion capability that throughput-efficient baselines cannot provide.

\noindent\textbf{Latency.}
We quantify the initial responsiveness of the evasion protocols using the Time to First Byte (TTFB) metric, defined as the total elapsed time from the client's initial TCP SYN until the first byte of application payload is received. To verify performance across realistic geographic deployments, we use the Linux kernel's \texttt{netem} module to inject fixed delays, configuring baseline RTTs of 20, 50, 100, and 200 ms.

Figure~\ref{fig:latency} visualizes the TTFB under varying network RTTs. \textit{Dummy} exhibits a linear dependency on network propagation, scaling from 80 ms to 800 ms as RTT increases, reflecting a strict $4\times$ RTT handshake overhead. \textit{UPGen} variants mirror this linear scaling, with heavy handshake configurations incurring larger round-trip multipliers ($6\times$ RTT). \emph{FlowPaint} exhibits a higher initial latency ($\approx$1.9s), dominated by the fixed inference cost of the generative model rather than network round-trips. 
However, as the baseline RTT increases tenfold (20 ms to 200 ms), \textit{Dummy} latency degrades by $10\times$, whereas \emph{FlowPaint} increases by only $\mathbf{1.28\times}$. This confirms that our system is compute-bound rather than network-bound: the 1.9s TTFB represents \textbf{a one-time startup cost}, as subsequent 1,024-packet blocks are processed in a fully pipelined fashion without recurring delays. Consequently, this overhead will naturally decrease with future advancements in hardware acceleration. Characterizing energy consumption, GPU occupancy, and performance across commodity hardware profiles remains an important direction for future work.

\bibliographystyle{ACM-Reference-Format}
\bibliography{sample-base}
\end{document}